\documentclass[letterpaper]{article} 
\usepackage{aaai18}  
\usepackage{times}  
\usepackage{helvet}  
\usepackage{courier}  
\usepackage{url}  
\usepackage{graphicx}  
\usepackage{capt-of}
\usepackage{comment}
\usepackage{multirow}
\usepackage{floatrow}
\usepackage{float}
\usepackage{booktabs}

\begin{document}
\title{Rating Reliability and Bias in News Articles: Does AI Assistance Help Everyone?}

\author{Benjamin D. Horne\textsuperscript{1}, Dorit Nevo\textsuperscript{1}, John O'Donovan\textsuperscript{2}, Jin-Hee Cho\textsuperscript{3}, and Sibel Adal{\i}\textsuperscript{1} \\
 Rensselaer Polytechnic Institute\textsuperscript{1}, University of California Santa Barbara\textsuperscript{2}, Virginia Polytechnic Institute and State University\textsuperscript{3}\\
 \{horneb, nevod, adalis\}@rpi.edu, jod@cs.ucsb.edu, jicho@vt.edu}

\maketitle
\begin{abstract} 
With the spread of false and misleading information in current
news, many algorithmic tools have been introduced with the aim of assessing bias and reliability
in written content. However, there has been little work exploring how effective these tools are at changing human perceptions of content. To this end, we conduct a study with 654 participants to understand if algorithmic assistance improves the accuracy of reliability and bias perceptions, and whether there is a difference in the effectiveness of the AI assistance for different types of news consumers. We find that AI assistance with feature-based explanations improves the accuracy of news perceptions. However, some consumers are helped more than others. Specifically, we find that participants who read and share news often on social media are worse at recognizing bias and reliability issues in news articles than those who do not, while frequent news readers and those familiar with politics perform much better. We discuss these differences and their implication to offer insights for future research.
\end{abstract}


\begin{table*}[htbp]
  \centering
  \hspace*{-0.2in}\begin{tabular}{ccc}
  \includegraphics[width=2.3in]{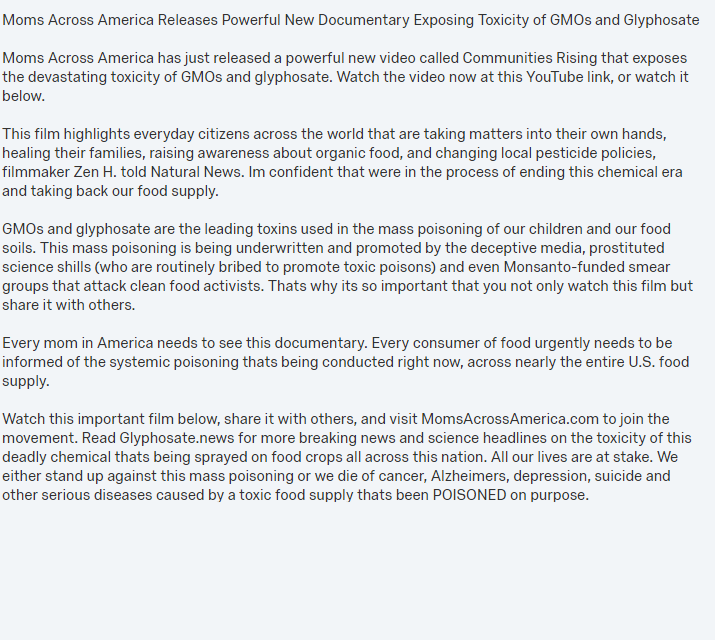} &
    \hspace*{-0.1in}\includegraphics[width=2.3in]{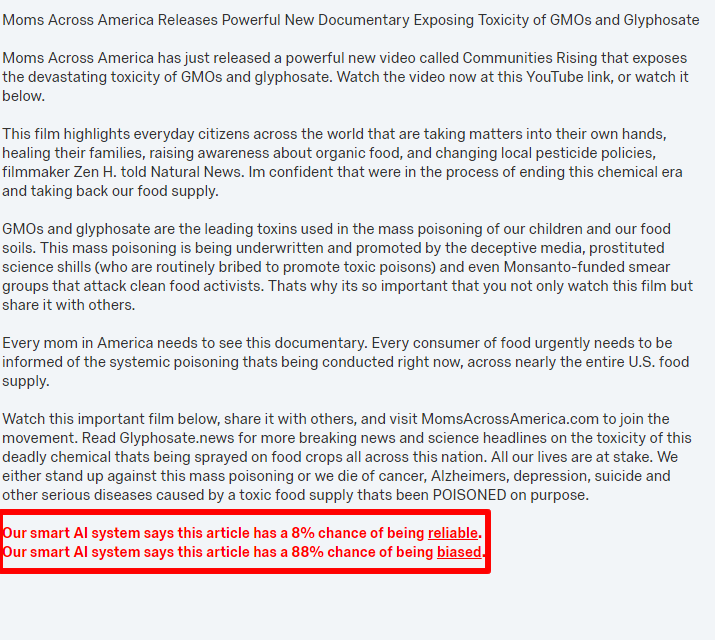} &
    \hspace*{-0.1in}\includegraphics[width=2.3in]{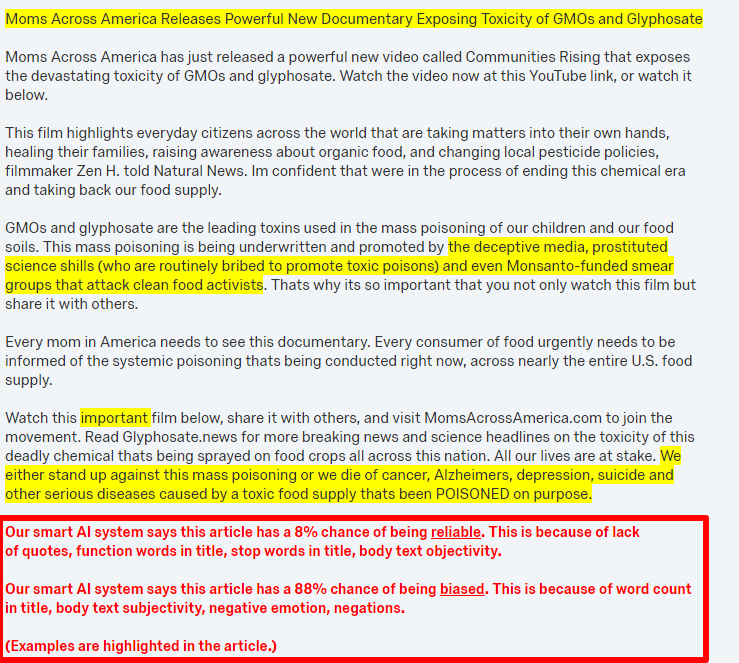}  \\
    (a) Text Only Condition & (b) AI Base Condition &  (c) AI Explanation Condition\\
  \end{tabular}
   \captionof{figure}{Screenshots of a news article in each condition. Note, the ``text objectivity'' and ``text subjectivity'' features are simply inverses of each other in the feature space. However, when automatic feature selection was done independently for the reliability model and for the bias model the differing features were selected. These features could be exchanged and not change the model.}
  \label{screenshots}
\end{table*}

\section{Introduction}
Today, false and misleading news are widespread and persistent~\cite{lewandowsky2012misinformation}. While both have long histories, they have become a recent focal point of researchers and practitioners due to the their surge in political news, and ultimately their negative impact on voting and public opinion worldwide~\cite{lazer2018science,lewandowsky2012misinformation}. This persistence of false information is aided by the structure of social media platforms, which boost passive news consumption and information overload~\cite{shu2017fake}. Information in the form of news articles is particularly prone to spreading misinformation (and disinformation) in this environment, as facts can be decontextualized in the headline to gain clicks, partial information can be reported in order to favor one side of an argument, or information can be completely fabricated to mimic news media content~\cite{chakraborty2016stop,lazer2018science}.

Due to the wide reach of false and misleading news, many automated methods have been introduced to counter their spread. These methods have focused on various aspects of information veracity, including bias~\cite{baly2018,budak2016fair,gentzkow2010drives,horne2018accessing}, reliability~\cite{baly2018,horne2017just,popat2016credibility,potthast2017stylometric,singhania20173han}, and source-level trustworthiness~\cite{pennycook2018crowdsourcing,swire2017processing}. In lab settings, many of these automatic methods have been shown to be highly accurate in detecting or approximating the integrity of information. For example, Popat et al. introduced a tool for assessing the credibility of individual claims using both content-based and source-based features ~\cite{popat2016credibility}. This tool achieved over 71\% accuracy (0.80 ROC AUC) on a large test set of true and false claims from Wikipedia. Similarly, Horne et al. built a tool for assessing the credibility of full news articles using content-based features~\cite{horne2018accessing}. This tool achieved 0.89 ROC AUC on a large weakly-labeled test set of news articles from various sources. Baly et al. achieved similarly high performance predicting source-level veracity using a mixture of features derived from content, Twitter, Wikipedia, and the source's web traffic~\cite{baly2018}.  

Despite these recent successes in automated news credibility, there has been little work exploring how these methods affect human subjects’ perception of the news. Focusing on human interactions with decision support algorithms, there has been numerous works on the general trust of users in algorithmic forecasts and on explaining algorithmic decisions to users~\cite{dietvorst2015algorithm,dietvorst2016overcoming,kleinberg2017human,kizilcec2016much,rader2018explanations}. However, these studies did not focus on the news context. We argue that it is very different to recommend an employee for promotion to a manager, or a weather forecast to a meteorologist, than it is to tell a user that the news she reads is misleading, false, or biased. This difference is because individuals often have strongly-held beliefs concerning the news~\cite{lazer2018science} and those individuals often use mental shortcuts or heuristics when assessing the news rather than deeper information processing~\cite{petty1986}. Furthermore, human interactions with decision support systems are often "pull"-based, where users seek a recommendation, while news credibility support may not be. Thus, it is necessary to better understand how algorithmically generated advice will be accepted in the unique context of misleading news content.

In this paper, we present an experimental study to fill this gap. Specifically, in our experiment individuals rate the credibility of news articles with or without algorithmic assistance. Our main goal is to understand if algorithmic assistance improves human decisions and whether there is a difference in performance with varying levels of algorithmic explanations and among different types of participants. Our paper contributes to the literature by providing a first look at human responses to AI assistance in the context of news veracity. We demonstrate using a real state-of-the-art AI assistant and real political news articles that AI assistance can improve human credibility perceptions. However, we also illustrate that some level of explanation is needed in order for the AI assistant to be compelling across different types of news. In addition, our results show that individual news consumption patterns impact the effectiveness of these tools. We demonstrate a significant difference in news perceptions between social media news consumers and expert news readers, as well as, their adherence to AI assistance. This set of results provides practitioners with insight into effective news assistant design. Finally, our work adds to the already rich literature on humans interaction with news of varying reliability and bias. 

\begin{table*}[!ht]
\centering
\fontsize{8.3}{11}\selectfont
  \hspace*{-0.0in}\begin{tabular}{|c|}
  \hline
\textbf{Unreliable article criteria}\\
\hline
1. Does the article have a misleading title (clickbait or makes claims not supported by the article)? \\
2. Does the article have no supporting evidence (missing quotes from witnesses, experts, or other reputable sources)?\\
3. Does the article have logical fallacies (claims are not supported by the evidence presented in the article)?\\
4. Does the article use overly emotional tone?\\
5. Is the article factually incorrect (claims made that can be shown as false or misrepresented)?\\
6. Does the article reference other unreliable sources (other sources that are known to produce false information)?\\
\hline
\hline
\textbf{Biased article criteria}\\
\hline
1. Does the article use overly emotional tone?\\
2. Does the article create a ``call to action'' (telling consumers what to think or to do)?\\
3. Does the article have framing bias (only reporting one side of the story)?\\
4. Does the article use subjective statements or opinion?\\
5. Is the title of the article one-sided (a headline that favors one side over another)?\\
\hline
\end{tabular}
  \caption{Criteria for expert article labeling, loosely developed from the criteria laid out in~\cite{zhang2018structured}. Each criteria is answered yes or no by the expert labeler.}
  \label{criteria}
\end{table*}

\section{Related Work}~\label{related} 
In order to grasp a more fine-grained view of news veracity, we explore two different constructs: the reliability of an article and the bias of an article. While highly related and often intertwined, reliability and bias are not the same concept~\cite{baly2018,horne2018accessing}, and can be perceived differently. Reliability concentrates on the factuality of reporting, whether an article contains true information, completely fabricated information, or misleading information. In contrast, bias is the imbalance of information from different sides of an issue or subjective opinions that can decontextualize truth~\cite{dellavigna2007fox,druckman2005impact,fico2004influence}. Both concepts can play a significant role in the spread of misinformation despite the different mechanisms used in each. Further, news articles can have varying levels of both reliability and bias, contributing to the articles overall veracity~\cite{baly2018,horne2018accessing}. Hence, we argue that these concepts need to be studied separately to gain a more complete understanding of when AI intervention is accepted or rejected in news decisions. 

It is clear there is a need for automated tools, as humans are prone to being misinformed by both unreliable information and biased information. Lewandowsky et al.~\cite{lewandowsky2012misinformation} examined cognitive factors that impact the spread of misinformation by humans. The authors show a large array of factors that influence news decisions, including (1) information that is compatible with what a person already believes can be seen as more credible, (2) stories that are coherent and compelling may be easier to believe, (3) if the information is from a source that is perceived to be credible, the information itself is perceived as credible, and (4) if others, particularly in a person's social circle, believe the information, it is seen as more credible. Another factor that can leave humans prone to misinformation is their information consumption patterns~\cite{del2016spreading,Bessi:2015hr}. Specifically, if a person is a part of a homogeneous and polarized group in which unverified information is abundant, they are more likely to share that unverified information (and hence cause information cascades)~\cite{del2016spreading}. In other words, if a person is in a tight-knit, false news spreading community, they may be acclimated to the style and structure of unreliable or hyper-partisan information, viewing it as more credible than it really is. Related to this notion is the literature on persuasion and credibility~\cite{petty1986}, which points out that news articles are interpreted in two separate levels: (1) at an emotional or heuristic level, to test whether information is relevant and is in sync with one's opinions, and (2) at a rational level, to test the veracity of the information. These levels are inseparable. Often heuristic methods are fast and require little cognitive effort and deeply impact the ``fact-checking'' aspect of information processing. Readers often reach a conclusion quickly using these heuristics and stop processing the information deeply, leaving them prone to being misinformed by false and misleading news. These mental shortcuts are often supported by the information overload and limited individual attention on social media platforms~\cite{mele2017combating}.

To this end, many automated methods have been introduced to counter the spread of false and misleading news. The majority of these methods use supervised, feature-based automatic detection, with features extracted from text content, the network in which the content spreads, and the user who spreads it. There are some more recent studies that explored deep learning and unsupervised methods~\cite{singhania20173han,shu2017fake}, and others that focused on crowd-based methods~\cite{pennycook2018crowdsourcing}; however, by far the most common and successful approach has been supervised content-based methods~\cite{baly2018,potthast2017stylometric,popat2016credibility,nakashole2014language,karduni2018can,zhang2018structured}. Many of these studies have shown high accuracy results on test sets of varying size, time-frame, and topic. While there are still some open questions about how well these methods work over longer periods of time and how well they generalize over changes in the news cycle, these works have shown promising results for AI assistance in news veracity tasks.

Despite much progress in both building these automated tools and understanding humans' deficiencies in news consumption, there has been little to no work to understand how the two work together. In general, there has been work on automated tools' impact on human decisions outside of the news context. For example, Dietvorst et al.~\cite{dietvorst2015algorithm} studied when humans choose the human forecaster over a statistical algorithm. The authors found that aversion of the automated tool increased as humans saw the algorithm perform, even if that algorithm had been shown to perform significantly better than the human. Dietvorst et al. explained that aversion occurs due to a quicker decrease in confidence in algorithmic forecasters over human forecasters when seeing the same mistake occur~\cite{dietvorst2015algorithm}. Hence, humans are more critical of algorithmic mistakes (mistakes by automated tools) than human mistakes. In a later study, Dietvorst et al. illustrated a decrease in algorithm aversion if the humans could slightly modify the forecasts~\cite{dietvorst2016overcoming}. This modification made the human participants feel more satisfied with the forecast and more likely to believe the algorithm was superior in predicting. Similar studies focused on the impact of generic algorithm explanations on the human's trust in the automated tool. Rader et al. performed an experiment of various ways to explain the Facebook News Feed algorithm~\cite{rader2018explanations}. They found that all explanation types tested helped participants become more aware of how the system works, but these explanations were less effective for evaluating correctness of the system's output. Thus the explanations were not necessarily useful in promoting trust in the algorithm. Kizilcec ran a similar experiment using a Massive Open Online Course (MOOC), and found that too much explanation of the automated tools' decisions can erode trust~\cite{kizilcec2016much}.

Broadly speaking, these works begin to explore the impact of algorithmic assistance on human decisions, but fail to address the context-specific nature of these tools and choices. As discussed, humans perceptions about the veracity of news can be rigid, strongly-held, and influenced by social pressure. These decisions are very different than decisions about the correctness of weather predictions, hiring a recommended job applicant, or trusting an algorithmic grading of a homework assignment. Our work begins to fill this context-specific gap. Specifically, we ask the following research questions: 
\begin{enumerate}
    \item (\textbf{Q1}) Does algorithmic assistance improve users' perceptions of news \textbf{reliability}?
    \item (\textbf{Q2}) Does algorithmic assistance improve users' perceptions of news \textbf{bias}?
    \item (\textbf{Q3}) Are there individual differences that impact the effectiveness of the algorithmic assistance in each case? (reliability perceptions and bias perceptions)
\end{enumerate}

\section{Experimental Design}
The study's objective is to understand how news consumers interact with algorithmic assistance. To this end, our experimental design includes three conditions. Under each condition, participants were asked to read news articles and rate these articles on their bias and reliability. The three experimental conditions were:
Condition 1: Article text only (\textit{text} condition)
Condition 2: Article text with AI assistance (\textit{AI base }condition)
Condition 3: Article text with AI assistance and explanations (\textit{AI explanation} condition)

In this section we describe first the construction of our article set. In line with our interest in understanding both bias and reliability we constructed an article set that focused on both constructs. Next, we explain in depth the AI assistant that was used in the study. The section concludes with a description of the study's respondents. 

\subsubsection{Article Set Construction}
We used a two-step approach to create our data set. First, we selected news sources that fall into three categories: (1) mainstream (typically assumed to be reliable and unbiased), (2) unreliable, and (3) biased. Specifically, unreliable sources are sources that have reported completely fabricated information in the past, and biased sources are sources that tend to report from a hyper-partisan point of view. We used two previously built lexicons to select these sources: the opensources lexicon (\url{www.opensources.co/}) and a hyper-partisan source lexicon from~\cite{pennycook2018crowdsourcing}. These definitions and source selections closely align with previous literature~\cite{baly2018,horne2018accessing,potthast2017stylometric}. The final list of sources used can be found in the Appendix (Table A\ref{dataset2sources}). 
Next, we extracted articles at random from the abovementioned sources, using a large 2017 news article data set~\cite{horne2018sampling}, and roughly balancing between the three categories of sources. We then followed with an expert rating approach, in which five experts (in this case four authors of the paper and one external communications expert) independently read and rated each article using a set of criteria (similar to those proposed in~\cite{zhang2018structured}), listed in Table~\ref{criteria}. Articles that were deemed unreliable (multiple criteria marked as ``yes'') by all raters (100\% agreement) were then defined as unreliable (UR, 13 articles) and maintained for this study. The same procedure was employed to define the remaining three types of articles: reliable (R, 11 articles), biased (B, 16 articles), and unbiased (UB, 9 articles). 

\subsection{AI Assistant}~\label{sec:ml_models}
To create an AI assistant, we use two state-of-the-art Random Forest classifiers developed in previously published work~\cite{horne2018accessing}.  Given a news article, these classifiers output the probability that the article is unreliable or biased, respectively. These classifiers are trained using a rich set of content-based features on a large set of political new articles. Some example features include, word usage, emotional tone, subjectivity, and sentence complexity. While there are numerous other recent methods for automatic news credibility classifications, we choose these content-based classifiers for several reasons: 
\begin{enumerate} 
\item The decisions made by content-based algorithms, specifically ones that use rich feature sets, can be explained.
\item Content based methods in general have been shown to be useful in news credibility task~\cite{baly2018,horne2017just,potthast2017stylometric,popat2016credibility}, and the vast majority of the literature on news credibility detection uses content-based methods in some form. 
\item These specific classifiers have been shown to be accurate in prior work, as well as, on our newly created data set. Specifically, according to~\cite{horne2018accessing}, these classifiers perform with ROC AUC scores above 0.90 for both reliability classification and bias classification. Furthermore, when testing the classifiers on our set of news articles, we found that the predictions matched our ground truth well. For articles labeled as reliable, the classifier reported 86.15\% reliability on average with a standard deviation of 10.62, while for articles labeled as unreliable the classifier report 17.18\% reliability on average with a standard deviation of 8.55. Similarly, for articles labeled as biased, the bias classifier reported 86.60\% bias on average with a standard deviation of 8.90, while for articles labeled as unbiased the classifier reported 26\% bias on average with a standard deviation of 8.10. No incorrect or uncertain probabilities (near 50\%) were reported. This high performance on our data set should be expected as it was created with strong ground truth.\end{enumerate}


Regardless of the method used in building the AI assistant, the task in this paper only requires our assistant to be accurate on our data set and that it's decisions can be clearly explained. Broader questions of generality or algorithm choice are left to another study.

 \begin{table*}[!ht]
\centering
\fontsize{8.8}{11}\selectfont
\begin{tabular}{| c | c | c | c |}
 \hline
& \textbf{Text Only} & \textbf{AI Base} & \textbf{AI Explanation}\\
 \hline
\textbf{N} & 217 & 211 & 226\\
 \hline
\textbf{Median Age Group} & 25-34 & 25-34 & 25-34 \\
 \hline
\textbf{Median Education Group} & 4 year college & 4 year college & 4 year college \\
 \hline
\textbf{Gender} & M: 100 F: 115 O: 2 &  M: 110 F: 100 O: 1 &  M: 100 F: 124 O: 2 \\
\hline
\textbf{Political Leaning} & VL:28 L:66 M:77 C:40 VC:6 & VL:30 L:57 M:85 C:28 VC:11 & VL:25 L:73 M:83 C:34 VC:11\\
 \hline
 \end{tabular}
 \caption{\label{demographics} Demographics for each treatment in our study, where both age and education are answered on 7 point scales.}
 \end{table*} 
 
 \begin{table*}[!htbp]
\centering
\fontsize{8.8}{11}\selectfont
\begin{tabular}{| c | c | c | c |}
 \hline
\textbf{Article Type} & \textbf{Text Only} & \textbf{AI Base} & \textbf{AI Explanation} \\
 \hline
\multirow{3}{*}{Reliable} & N: 217 & N: 211 & N: 226 \\
& Average rating: 6.64 & Average rating: 7.34 & Average rating: 7.10\\
& Standard Dev: 2.15 & Standard Dev: 2.00 & Standard Dev: 1.86 \\
 \hline
\multirow{3}{*}{Unreliable} & N: 217 & N: 211 & N: 226 \\
& Average rating: 5.01 & Average rating: 4.79 & Average rating: 3.84\\
& Standard Dev: 2.26 & Standard Dev: 2.52 & Standard Dev: 2.27 \\
 \hline
\multirow{3}{*}{Unbiased} & N: 214 & N: 202 & N: 225 \\
& Average rating: 4.37 & Average rating: 4.42 & Average rating: 3.89\\
& Standard Dev: 2.59 & Standard Dev: 2.70 & Standard Dev: 2.05 \\
 \hline
\multirow{3}{*}{Biased} & N: 217 & N: 211 & N: 226 \\
& Average rating: 6.58 & Average rating: 6.97 & Average rating: 7.15\\
& Standard Dev: 2.28 & Standard Dev: 2.34 & Standard Dev: 1.86 \\
 \hline
 \end{tabular}
 \caption{\label{results1} Summary statistics for each condition and article ground truth. Note, ratings closer to 10 are better for reliable articles and biased articles, while ratings closer to 1 are better for unreliable articles and unbiased articles. One-way ANOVA results for each condition and article type can be found in text below.}
 \end{table*}

\subsection{Human Experiment on Amazon Mechanical Turk}~\label{mturk}
Using this data set and AI assistant, we conducted a randomized between-subjects study on three conditions (\textit{text} condition; \textit{AI base} condition; and \textit{AI explanation} condition). In each condition, participants (Turkers) rated articles on their bias and reliability. Participants were asked to rate articles' reliability on a 10-point scale ranging from `completely unreliable' (1) to `completely reliable' (10). They were similarly asked to rate articles' bias on a 10-point scale ranging from `completely unbiased' (1) to `completely biased' (10). After each rating, participants were also asked to comment on why they rated the article this way, in order to qualitatively capture any features or thoughts the participant used in their assessment.


In the \textit{text} only condition, participants were only given the article text, including both the body text and the headline text. No additional information such as the journalist or source that wrote the article is provided. Each participant evaluated between three to five randomly assigned articles, where three to five is chosen based on how long we wanted the user experience to be. If a participant is given more than one article of the same ground truth (R, UR, UB, B), we only keep the first rating to avoid repeated measures. 


In the \textit{AI base} condition, we introduced our AI assistant (discussed in Section 3.2), which provides the user with a predicted probability of each article being reliable or biased along with the article text. We displayed this prediction at the bottom of each article stating ``Our smart AI system says this article has a X\% chance of being (reliable or bias)'' in bold red font. All other parts were formatted exactly as the text only condition. 

In the \textit{AI explanation} condition, participants were again shown an AI prediction with the article text, but in addition a feature-based explanation of the prediction is shown. Specifically, we showed the top four most important content features for each predicted class and highlighted four to eight examples of them in the article. These features were easily extracted from our AI assistant as the model is built using decision trees. Specifically, we computed the mean decrease impurity of a feature averaged over all trees in the ensemble model~\cite{breiman1984classification,pedregosa2011scikit}. This provides us with a feature importance ranking for each article. Further, since the features are based on content, they are easily interpretable. All other parts were formatted exactly as the text only condition. Figure~\ref{screenshots} presents screenshots of all three conditions. 

While, there are other methods to explaining the AI decisions, such as example-based explanations, we focus on one type of explanation method to simplify the analysis. Varying the presentations of information are left for future work.

\subsubsection{Participants}~\label{partic}
Participants were workers on Amazon Mechanical Turk (AMT). Based on recommendations from the literature on using AMT for research, a filter was applied to ensure that participants had successfully completed at least 50 tasks in the past~\cite{dennis2018mturk}. Responses were provided using a Qualtrics survey.

In addition to the experimental task described above, participants were also asked about demographics information, as well as their news consumption habits. Specific questions on individual differences were:
\begin{enumerate}
\item How familiar are you with US politics? (5-point scale ranging from ``not at all'' to ``extremely'')
\item How often do you read news? (4-point scale ranging from ``never'' to ``multiple times a day'')
\item What is the primary way you get news? (``social media'', ``news websites'', ``TV'', ``newspaper'')
\item When you use social media, how often do you share news? (4-point scale ranging from ``never'' to ``always'')
\item To what extent do you trust news coming from mainstream media? (``don't trust''; ``do trust'')
\item To what extent do you trust news coming from your social contacts? (``don't trust''; ``do trust'')
\item What is your political leaning? (5-point scale ranging from ``very liberal'' to ``very conservative'')\footnote{We also asked three policy-based political questions to validate the self-reported political leaning question. No unexpected differences between policy and self-selected leaning were found.}
\end{enumerate}

This information was used in our analysis of the results to identify contingencies in acceptance of the AI assistance, as we explain in the next section. 
Finally, to ensure reliability in responses (for example - to avoid the problem of bots used on AMT or users clicking through survey with little effort) we included one check question in each survey  and one simple reading comprehension question for each article. Data were cleaned to ensure all responses were of sufficient quality. Similar attention check filtering processes are used in Amazon Mechanical Turk studies throughout the literature~\cite{kittur2008crowdsourcing,knijnenburg2015evaluating} and have been recommended in meta-analysis~\cite{dennis2018mturk}. 


Table~\ref{demographics} describes the final groups of respondents as well as their demographic information.


\section{Results}
We conducted four one-way ANOVA tests corresponding with the four ground truths of interest in this study (Reliable, UnReliable, Biased, and UnBiased articles). For each test we used respondents' rating as the dependent variable, and the experimental condition as the independent variable. Table~\ref{results1} provides descriptive statistics for each condition, and the following text summarizes the results of the ANOVA tests. 
 
 \begin{table*}[h]
\centering
\fontsize{8.8}{11}\selectfont
\begin{tabular}{| c | c || c |}
\hline
\textbf{Article Type} &\multicolumn{1}{|c||}{\textbf{Political Familiarity}} & \multicolumn{1}{|c|}{\textbf{Reading Frequency}}\\
\hline
\multirow{2}{*}{Reliable} & No significant interaction & No significant interaction \\
& $F_{condition} = 7.66$**; $F_{familiarity} = 15.14$** & $F_{condition} = 6.69$**; $F_{reading} = 15.87$**\\
\hline
\multirow{2}{*}{Unreliable} & No significant interaction & No significant interaction\\
& $F_{condition} = 31.6$** &  $F_{condition} = 32.12$**\\
\hline
\multirow{2}{*}{Unbiased} & No significant interaction & No significant interaction\\
& $F_{condition} = 4.61$**& $F_{condition} = 4.34$*; $F_{reading} = 5.29$*\\
\hline
\multirow{2}{*}{Biased} & No significant interaction & No significant interaction\\
& $F_{condition} = 9.12$**; $F_{familiarity} = 10.65$** & $F_{condition} = 8.11$**\\
\hline
\hline
&\multicolumn{1}{|c||}{\textbf{Where News is Read}} & \multicolumn{1}{|c|}{\textbf{Sharing Frequency}}\\
\hline
\multirow{2}{*}{Reliable} & No significant interaction & No significant interaction\\
& $F_{condition} = 6.38$* & $F_{condition} = 7.23$**\\
\hline
\multirow{2}{*}{Unreliable} & No significant interaction & No significant interaction\\
& $F_{condition} = 31.95$** &  $F_{condition} = 28.27$** $F_{sharing} = 20.97$**\\
\hline
\multirow{2}{*}{Unbiased} & No significant interaction & No significant interaction\\
& $F_{condition} = 3.96$*& $F_{sharing} = 29.03$**\\
\hline
\multirow{2}{*}{Biased} & No significant interaction & No significant interaction \\
& $F_{condition} = 8.00$**& $F_{condition} = 7.27$**\\
\hline
\hline
&\multicolumn{1}{|c||}{\textbf{Trust in Mainstream Media}} & \multicolumn{1}{|c|}{\textbf{Trust in Social Contacts}}\\
\hline
\multirow{2}{*}{Reliable} & Significant interaction ($F_{interaction} = 4.68$*) & No significant interaction\\
& $F_{condition} = 5.33$*; $F_{mainstream} = 35.37$** & $F_{condition} = 6.49$*\\
\hline
\multirow{2}{*}{Unreliable} & Significant interaction ($F_{interaction} = 4.16$*) & No significant interaction\\
& $F_{condition} = 33.29$** & $F_{condition} = 31.93$** $F_{socialtrust} = 13.06$**\\
\hline
\multirow{2}{*}{Unbiased} & No significant interaction & No significant interaction\\
& $F_{condition} = 4.69$*; $F_{mainstream} = 3.88$*& $F_{condition} = 4.00$*\\
\hline
\multirow{2}{*}{Biased} & No significant interaction & No significant interaction\\
& $F_{condition} = 7.64$** & $F_{condition} = 7.92$** \\
 \hline
  \bottomrule
\addlinespace[1ex]
\multicolumn{2}{l}{\textsuperscript{**}$p<0.01$, 
  \textsuperscript{*}$p<0.05$}
 \end{tabular}
 \caption{\label{results2} Two-way ANOVA results for each individual difference measure. For measures that had a significant interaction, refer to Figure~\ref{interact}. Note, two-way ANOVA examines the influence between two independent variables (AI conditions and individual difference) and a dependent variable (ratings). If there is an interaction, the effect of one factor depends on the other factor.}
 \end{table*} 
 
 \begin{table}[h]
\centering
\fontsize{8.8}{11}\selectfont
\begin{tabular}{| c | c |}
\hline
 \textbf{Article Type} &\multicolumn{1}{|c|}{\textbf{Political Leaning}}\\
\hline
\multirow{2}{*}{Reliable} & No significant interaction \\
& $F_{condition} = 7.72$**; $F_{leaning} = 6.61$**  \\
\hline
\multirow{2}{*}{Unreliable} & No significant interaction   \\
& $F_{condition} = 32.82$** \\
\hline
\multirow{2}{*}{Unbiased} & No significant interaction \\
& \\
\hline
\multirow{2}{*}{Biased} & No significant interaction \\
& $F_{condition} = 12.55$** \\
 \hline
  \bottomrule
\addlinespace[1ex]
\multicolumn{2}{l}{\textsuperscript{**}$p<0.01$, 
  \textsuperscript{*}$p<0.05$}
 \end{tabular}
 \caption{\label{results2} Two-way ANOVA results for political leaning measure. Refer to Table~\ref{results2} for other individual measures.}
 \end{table} 
 
 \begin{table}[h]
 \centering
\fontsize{8.8}{11}\selectfont
\begin{tabular}{| c | c | c | c |}
\hline
\multicolumn{4}{|c|}{\textbf{Text Only}} \\
\hline
& \textbf{Liberal} & \textbf{Moderate} & \textbf{Conservative} \\
\hline
\textbf{Left Articles} & 6.82 & 6.07 & 6.84\\
\hline
\textbf{Right Articles} & 6.49 & 6.19 & 6.30\\
\hline
\hline
\multicolumn{4}{|c|}{\textbf{AI Base}} \\
\hline
& \textbf{Liberal} & \textbf{Moderate} & \textbf{Conservative} \\
\hline
\textbf{Left Articles} & 6.90 & 6.69 & 7.95\\
\hline
\textbf{Right Articles} & 6.97 & 6.75 & 7.37\\
\hline
\hline
\multicolumn{4}{|c|}{\textbf{AI Explanation}} \\
\hline
& \textbf{Liberal} & \textbf{Moderate} & \textbf{Conservative} \\
\hline
\textbf{Left Articles} & 7.39 & 7.31 & 6.91\\
\hline
\textbf{Right Articles} & 6.72 & 6.96 & 6.54\\
\hline
 \end{tabular}
 \caption{\label{politics} Average rating of participants when rating \textbf{biased articles}, broken down by article and participant leaning. Since there were very few participants on the extreme-ends (very liberal or very conservative), we combine the liberal and conservative groups.}
 \end{table} 
 
 \begin{table}[ht]
\centering
\begin{tabular}{c}
(a) Reliable Ground Truth \\
\includegraphics[width=2.2in]{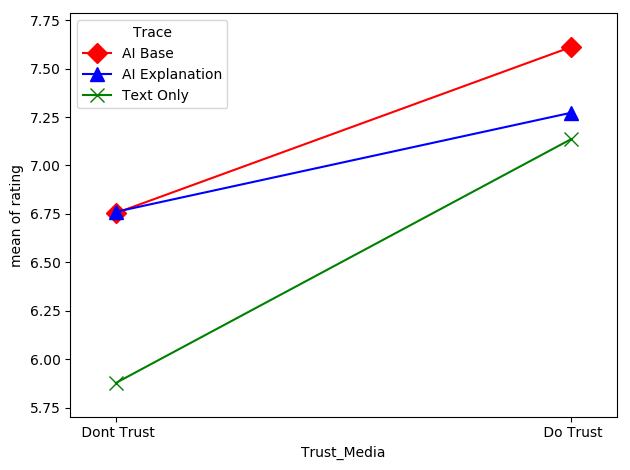} \\
(b) Unreliable Ground Truth \\
\includegraphics[width=2.2in]{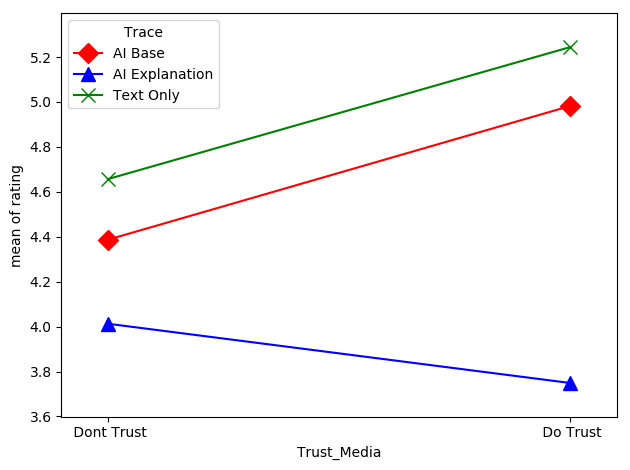}  \\
\end{tabular}
\captionof{figure}{Interaction plots for \textbf{Trust in Mainstream Media}}
\label{interact}
\end{table}
 
For the reliable (R) articles, the ANOVA showed a significant difference in group means (F=7.8129, sig. 0.0004), with the post hoc test indicating the different means were between the text only and AI base conditions as well as the text only and AI explanation conditions. No significant difference was found between the two AI conditions.
 
For the unreliable (UR) articles, the ANOVA showed a significant difference in group means (F=18.1541, sig. 0.000), with the post hoc test indicating the different means were between the text only and AI explanation as well as the two different AI explanation conditions. No significant difference was found between text only and AI base conditions. 

For the unbiased (UB) articles, the ANOVA and post hoc tests did not show a significant difference in group means (F=2.9945, sig. 0.051). 

Finally, for the biased (B) articles, the ANOVA showed a significant difference in group means (F=4.1947, sig. 0.015), with the post hoc test indicating the different means were between the text only and AI explanation conditions.  No significant difference was found between the text only and AI based conditions nor between the two AI conditions. 

\subsection{Analysis of individual differences}
To delve deeper into the contingencies that may impact individuals' perceptions of news articles, as well acceptance of AI assistance, we followed with a set of two-way ANOVA tests. Each test used the same dependent variable and experimental conditions as before (the individual ratings under the three conditions) but added a second factor capturing differences in news consumption. The questions concerning these individual differences were described in Section~\ref{mturk}. We conducted a total of 24 two-way ANOVA tests (the four article categories (R,UR,UB,B) and the six news consumption questions). Note, two-way ANOVA examines the influence between two independent variables (AI conditions and individual difference) and a dependent variable (ratings). If there is an interaction, the effect of one independent variable depends on the other independent variable. This significant interaction means we cannot clearly interpret the significance of the independent variables on the dependent variable. In other words, if we have no significant interaction, but a significant difference in the independent variables, we can say something about the impact of those factors on the dependent variable. We present the results below in Table~\ref{results2}. 
 
 \section{Discussion} 
 \paragraph{AI Assistance improves human perceptions about reliability and bias in news articles, but explanations are needed.}
 In general, we find that AI assistance does improve participants' perceptions of reliability and bias in news articles. However, this improvement depends on the level of explanation used by the tool. Specifically, we find that both the \textit{AI Base} condition and the \textit{AI Explanation} condition significantly improve participants' ratings on reliable articles, but only the \textit{AI Explanation} condition improves participants' ratings on unreliable articles and biased articles. Furthermore, in our two-way ANOVA analysis we find that the AI assistant helps every individual difference group significantly, with exception of those who share news often on social media. We explore this exception further below. 
 
 \paragraph{Political familiarity and reading frequency help participants judge when information is reliable.}
 Our two-way ANOVA results show that both how familiar a participant is with politics and how often a participant reads news significantly improve the judgment of reliable articles (see Table~\ref{results2}). Specifically, those who said they are very familiar with politics had an average rating of 6.6 in the text only condition, while those who said they are not at all familiar with politics had an average rating of 6.3. This difference increases with AI assistance, where those very familiar rated reliable articles 7.7 on average and those not familiar rated articles 6.6 on average in the \textit{AI Base} condition. These large differences in ratings also hold true for the \textit{AI Explanation} condition. Interestingly, these differences do not hold when participants rated unreliable articles. 
 
 Similarly, those who read news multiple times a day had an average rating of 6.7 while those who never read the news had an average rating of 6.4 in the \textit{Text} condition. Again, this difference is increased in the AI conditions. Those who read news multiple times a day rated reliable articles 7.6 on average and those who never read news rated reliable articles 5.6 on average, which is even worse than they did without the AI assistant. 
 
 
 \paragraph{Those who trust news from social contacts and share news often on social media perceive unreliable articles as more reliable.}
 Our two-way ANOVA results also show that trust in news from ones social contacts and how often a participant shares news on social media significantly impacts ratings of unreliable articles (see Table~\ref{results2}). On average, those who trust news from social contacts rated unreliable articles 5.25, while those who do not trust rated articles 4.50 in the \textit{Text} condition. While both groups improve significantly in the \textit{AI Explanation} condition, the difference in performance still exists, with those who trust rating unreliable articles 4.0 on average and those who do not trust rating unreliable articles 3.50 on average. Remember, a lower rating for unreliable articles is more accurate compared with our ground truth, therefore, those who do not trust in news shared by their social contacts do better without AI and with AI than those who trust. Even more significantly, those who reported sharing news on social media often did much worse at rating unreliable articles (average rating without AI 6.25) than those who reported they never share news on social media (average rating without AI 4.5). Again, AI with explanation significantly helps both groups, but differences in performance still exist, with those who often share rating unreliable articles 4.25 on average and those who never share rating unreliable articles 3.75 on average.
 
 This set of results suggests that heavy social media users perceive unreliable news articles as more reliable than they actually are. Previous work shows that fake news has very different style and structure than traditional news~\cite{horne2017just} and that repeated exposure to false news is correlated with believing and sharing false news in the future~\cite{lewandowsky2012misinformation,Bessi:2015hg,del2016spreading}. Furthermore, we know that during recent U.S. elections, fake news stories were more widely shared than mainstream news stories on Facebook~\cite{allcott2017social}, illustrating widespread exposure to the style and structure of unreliable news articles. While we do not have knowledge of the specific online communities or platforms that our study participants read and share in, this finding could be due to widespread repeated exposure to unreliable structured news articles on social media. Further study is needed to explain these results conclusively. 
 
 
 \paragraph{Political familiarity and trust in mainstream media help perceptions of bias in news articles.}
 Similar to our findings with reliable news articles, we find that participants who are more familiar with politics and trust mainstream media are better at recognizing bias in news articles. Without AI assistance, those who said they are very familiar with politics rated biased articles 6.75 on average, while those who said they are not at all familiar with politics rated biased articles 5.75 on average (keep in mind that higher ratings are better for bias ground truth). In the \textit{AI Explanation} condition, both groups improve (7.25 average for those very familiar and 6.50 average for those who are not familiar). Additionally, those who said they trust the mainstream media were better at recognizing bias without AI assistance, rating biased articles 6.8 on average, than those who do not trust mainstream media, rating biased articles 6.25 on average. These differences effectively go away in both AI conditions.
 
 \paragraph{Feature-based AI assistance is helpful in pointing out bias, but not necessarily the lack of bias in a news article.} 
 In our one-way ANOVA results, we see significant improvement in participants recognition of bias in news articles, but not recognition of the lack of bias in news articles. At a high-level, this result makes sense, as our feature-based classifiers can highlight biased statements, such as subjective language or emotional tone, but may not be able to clearly highlight examples of being unbiased, particularly without specific knowledge of the issue being discussed. Despite there being no significant improvement in our one-way ANOVA results, we do see significant improvement in recognizing lack of bias in our two-way ANOVA results. All individual groups were significantly affected by the AI assistant conditions when rating unbiased articles, except for the social news sharing group. Looking closer at the means of each type of participant in this group, we see that those participants who share news on social media 'some of the time' (as opposed to 'never' or 'most of the time') are the only group that improves with AI assistance. However, in all three conditions, those who never share news on social media are better at recognizing lack of bias in news articles than those who share news on social media often. 

\paragraph{Political leaning has little impact on rating reliability and bias.} In Table~\ref{results2}, we show the two-way ANOVA results for the political leaning of participants. Surprisingly, we find that the political leaning of participants does not have a strong impact on article ratings overall. We only find significant differences in ratings for reliable articles ($F_{leaning} = 6.61$**). When looking at a Tukey post hoc test, we see the only group that differs is the ``very liberal'' group, which seems to rate reliable articles as more reliable. This difference does not exist for the other types of articles. 

In Table~\ref{politics}, we break down the average rating of participants in each political ideology with the political leaning of the biased articles rated. We can see in each condition, the average difference between rating an article of the same ideological leaning and of the differing ideological leaning is very small, and sometimes non-existent. For example, that both liberal and conservative participants rated left leaning articles as slightly more biased than right leaning articles. Similarly, in the AI Base condition, conservatives rated both left and right leaning articles as more biased than liberals. Thus, article ideology seems to be little to no impact on ratings. It should be noted that the distribution of political leanings is slightly skewed towards ``liberal'' with most participants being either ``moderate'' or ``liberal'' (see Table~\ref{demographics}). This slight skew is expected based on previous Mturk studies, but may influence the results we find concerning political ideology. Furthermore, this set of articles is not chosen based on strong ideological issues, but on general bias. It may be that if both right and left biased articles were selected to be on the same topic from the same time frame, ideology may come into play. However, this was not the goal of the study. 
 
 \section{Conclusion and Limitations}
 In conclusion, this study is the first in exploring the effectiveness of AI assistance in news credibility perceptions. We presented an experimental study in which humans rated the reliability and bias of real news articles with varying levels of assistance from a state-of-the-art decision support tool. We found that AI assistance with feature-based explanations significantly improved perceptions of reliability and bias. However, these improvements differ between different types of news consumers. Some participants tended to do well on their own, particularly if they reported to have high expertise through reading news frequently or political familiarity. In comparison, participants who used social media heavily showed negative results, perceiving unreliable articles as more reliable. While both groups improved significantly, those who share and read news on social media never perform as well as those who do not. For further insight, we provide qualitative analysis of the users' explanation of their ratings in the Appendix. 
 
Our results suggest that AI tools for news credibility can be most effective if they explain how the decisions are made rather than act as a black box. Further, while our results suggest AI assistance will help everyone to some extent, tools may be even more effective if they are tailored to individual differences. For example, our study only covered one type of automated assistant and explanation type (namely feature-based explanation), but it may be more effective to leverage a frequent social media users' friends to change their belief about the veracity of an article, similar to previously proposed crowd sourced methods~\cite{pennycook2018crowdsourcing}. On the other hand, those who more actively read news may be most helped by feature-based explanations, as our study used. Similarly, our study shows that reliability and bias are judged differently, and those judgments are influenced differently by the AI assistance. Hence, it may be useful to tailor an AI assistance to explain bias decisions differently than reliability decisions. 

More research is certainly required to properly assess the dynamics of trust placed in the AI assistant, and to assess levels of adherence to its predictions, in cases of agreement and disagreement with prior belief, and the knock-on effects on repeated interactions with the advice giver overtime. Additionally, other news consumption environments should be tested. Specifically, participants in our study are forced to read and at least minimally comprehend each news article. This consumption is different than the often passive consumption of news on social media, where users may only read the title or skim a news article. Thus, while our study provides a first step in understanding the effectiveness of news assistance, it is only looking at active consumption interactions, not passive consumption interactions. Although, it is certainly true that news consumption can happen actively on social media, further study should take place to also understand news assistance when consumption is passive. Another important limitation of this study is the explanation method used. Our study focused on one very simple explanation method, but there are many more ways to present or explain the results of an algorithm. It may be the case that participant's adherence to algorithm advice improves or worsens with other explanation methods. In the future, we want to explore and compare other explanation methods such as example-based explanations. Lastly, in our participant instructions, we did not explicitly define reliability or bias, but rather left it up to the participants interpretation. It is possible that not having a briefing on these concepts created some noise in our response data. In future work, we will address these concerns.

\section{Acknowledgements}
\small{We would like to thank Tamar Gordon for being our external expert in the article labeling task. We would also like to thank the reviewers for their many helpful suggestions for this paper.}
 

\bibliographystyle{aaai}
\bibliography{references}

\setcounter{figure}{0}
\setcounter{table}{0}
\section{Appendix}
\begin{table}[!htbp]
\centering
\fontsize{8}{9}\selectfont
\begin{tabular}{| c | c | c |}
\hline
\textbf{Mainstream sources} & \textbf{Unreliable sources} & \textbf{Hyper-partisan sources}\\
\hline
Associated Press & Infowars & Brietbart \\ 
PBS & Liberty News & Young Cons \\
NPR & Natural News & RedState  \\
CBS & Alt Media Syndicate & TheBlaze \\
USA Today & DC Clothesline & Politicus USA \\
BBC & Newslo & Bipartisan Report \\
The Guardian & Freedom Daily  & Occupy Democrats \\
 & Daily Buzz Live & Daily Kos   \\
 & Intellihub & Shareblue \\
\hline
\end{tabular} 
\caption{\small{Sources used in stage one of article construction. Note, only US stories were selected from BBC.}}
\label{dataset2sources}
\end{table}

\begin{table}[h]
    \centering
\fontsize{8}{9}\selectfont
\noindent\begin{tabular}{p{3in}} \hline
{\bf Writing Style}   \\
It uses a lot of opinion statements, and not a lot of evidence\\
Written in a convoluted style \\
The informal, accusatory, aggressive tone of the writing. \\
This is a made-up news article, because it doesn't follow Associated Press style for capitalization. \\
The headline is totally unprofessional.\\ 
The story has a lot of grammatical errors \\
A lot of negative emotions, Some language seems sensational \\
It doesn't have emotional flash points or inflammatory language. \\
Seems to be coherent and in order \\ \hline
{\bf AI Advice} \\
I based it on the AI system since I know nothing about this. \\
The AI system rating lead me to think that this is unreliable article. \\
It provides updates to previously reported news, stating facts and the smart AI system gave a 95\% chance \\
I'm going with the AI on this one\\
Strong AI rating.\\ \hline
{\bf Journalistic Features} \\
Didn't really have cites to back this up \\
 Includes a non sequitur \\
 Because it uses unnamed sources to make its statement \\
 It addressed both sides of the question without seeming to take sides. \\
\hline
{\bf Trust} \\
 Can never be 100\% sure if news is real or fake these days \\
 The FBI can't be trusted.  \\ 
\hline
{\bf Other Heuristics} \\
Clearly biased article written by an angry feminist \\
 It seems logical that these events happened. \\ \hline
 \end{tabular}
 \caption{\small{Example comments from users on their article ratings.}}
    \label{tab:freetext}
\end{table}




\end{document}